\documentstyle[11pt]{article}
\title{
\vspace{-1.in} \hfill {\small\rm UTPT-94-06}
\\~\\~\\
A New  Paradigm for the Fermion Generations}
\author{George Triantaphyllou
\thanks{e-mail:
george$@$medb.physics.utoronto.ca}\\
\\
Department of Physics, University of Toronto
\\ Toronto, ON M5S 1A7, Canada}
\begin{document}
\setlength{\baselineskip}{24pt}
\maketitle
\begin{abstract}
%\abstract
{\noindent
A new mechanism is proposed to explain the appearance
 of the three known fermion generations in a natural way.
 The underlying idea is based
 on the discreteness of the spectrum of solutions
 of the gap equation appearing in
 models of dynamical chiral symmetry breaking.
  Within such a framework, the number of
  parameters needed
 to describe the experimentally observed fermion spectrum
  is drastically
 reduced. The phenomenological consequences of such
  a mechanism are carefully
 discussed, in order to explore its viability.
}
 \vspace{1.in}
\end{abstract}
\begin{center}
{\it Published in Helvetica Physica Acta, {\bf 67}, No. 6 (October 1994)}
\end{center}
 \vspace{1.in}
\pagebreak

\section {Introduction}
After many years of theoretical investigations
 focused on
possible mechanisms responsible for the electroweak
 symmetry
breaking and the generation of
fermion masses, the puzzle still remains unsolved.
Even though the minimal
Higgs mechanism seems to be consistent with current
 experimental data \cite{LEP},
it remains
conceptually unsatisfactory, due to the excessive
fine tuning that must
be applied to the Higgs coupling in order to keep the renormalized
 Higgs mass reasonably
close to the weak scale.

The two presently most prominent alternatives to
the Higgs mechanism are
supersymmetry and technicolor, even though they could be far from
providing the correct mechanism hidden behind the Higgs sector.
They tackle the problem of weak $SU(2)_{L}$ breaking
in quite different ways. Nevertheless, they have a
 common feature: they
associate each fermion mass
with a different coupling, a Higgs coupling when it comes
 to supersymmetry, and an
extended-technicolor coupling when it comes to
technicolor. This loads
these two  theories with too many parameters,
 making them unnatural in that
respect. Model-builders for both theories
 typically hide this problem
 ``under the rug", placing
 the natural origins of these couplings,
 and their effects to the fermion family replication,
 to
  unknown physics at much
higher energies.

There have been attempts to
reduce the number of free parameters by
predicting the number of fermion generations
within the framework of grand-unified theories (supersymmetric or not)
\cite{GUT},
compositeness models \cite{compos}, string theory \cite{string},
as well as many other ideas \cite{crazy}. In
most of
 these cases, however, what is naturally derived
 is the number of generations,
and not the particular scale of the fermion masses in these generations.
Moreover, in the case of GUT, the number of generations is presented as
a constraint imposed by phenomenology, with no fundamental explanation.
In supersymmetry, there has been
 a recent attempt to reduce
considerably the number of free parameters
 entering
the fermion spectrum of the theory \cite{Dimopoulos}.
 Although such
an attempt seems to be headed towards the correct
 direction, it is still
plagued with a draw-back: the number of fermion
generations,
and the hierarchy between them, is again
introduced {\em ad hoc}, with no underlying
mechanism presented as responsible
for it.

Moreover, recent
 extended-technicolor models, in their
attempt to decrease the technicolor
 contribution to
the $\Delta\rho$ parameter, introduce
 even more parameters, associating
with each ordinary fermion not only each
own extended-technicolor coupling, but also each
own extended-technicolor scale \cite{Einh}.
 The idea of multiple scales
appeared in the early days of
technicolor theories, in the context of ``tumbling" \cite{tumble},
 which still remains a popular idea \cite{ApTer}.
In all these studies, however, the hierarchy of the fermion
generation scales is put in
by hand, without being presented within a theoretical framework
that would justify their magnitude or multiplicity.
 General model building
considerations might give one an idea of what these scales should be,
but it is our feeling that they do not tackle the problem in the most
fundamental way.

Such a situation is obviously far from
desirable for a natural theory.
Especially for technicolor, since one
 introduces additional degrees of freedom, in the form of a new
non-abelian
gauge group and families of new,
presently unobserved, fermions,
 one would
expect a much smaller number of parameters
 needed to explain the ordinary fermion
spectrum, without having to resort to fermion compositeness or
``barock" models.

The present paper  proposes a mechanism that
 could potentially
explain how the known fermions acquire their
 masses and they are at the same time
placed in distinct generations. The analysis is  done
within the framework of technicolor theories,
 since it is not presently
clear to the
author how a mechanism based on similar principles
could be at works in supersymmetric theories.
The method relies heavily on
the Schwinger-Dyson gap equations, solved by
using the typical
assumptions and approximations that are used in technicolor models
\cite{TC}.
The idea central to the development of the paper
is that the new physics,
 introduced by the extended-technicolor
 interactions at a scale $\Lambda_{ETC}$,
  the scale at which the extended-technicolor group breaks,
act as an effective cut-off to the integral
 gap equations which give the
self-energies of the fermions. This provides the theory with
a discrete
 spectrum of solutions.

  An
 attempt is then made to associate
the first three solutions of this discrete
 spectrum with the three fermion
generations, and to find physical arguments
 that would allow the
truncation of this spectrum beyond the third
solution, since present
electroweak-experiment data \cite{LEP} constrain the
fermion generations to three. The method gives results that are
directly testable, since it produces explicit
values for the order of magnitude of
masses and scales
that can be easily discarded if they grossly contradict phenomenology.
At first sight, it is not apparent to us how such a mechanism
could tell us something precise about the CKM matrix, so we are
not going to address that problem here. Moreover, it
should be noted that, because the equations thar appear in
this analysis are very difficult to solve exactly, the
results presented here try to sketch the qualitative
features of the mechanism, with no ambition
for providing quantitatively reliable results.

This is how this work is organized: At first,
 the general setting of
technicolor theories is overviewed, including
 extended-technicolor interactions.
Then, the  role that the integral Schwinger-Dyson
equations play in
the dynamical symmetry
breaking is analysed in a somewhat novel way,
and it is made clear how their
properties could help us solve the fermion-spectrum
 puzzle.
The results of this analysis show that a certain
combination of the physical parameters of some models
 obey a quantization
condition. The next section  tries to motivate such
 a quantization condition physically in various ways,
 and to check whether the predictions of this mechanism
 are consistent with current phenomenology.
 The final section
 summarizes the conclusions drawn by this analysis, and attempts to
 test the naturalness and viability of the proposed mechanism.

\section{ Mass generation in technicolor}

As was mentioned in the previous section, the Higgs mechanism,
in trying to explain the breaking of the $SU(2)_{L}$ gauge group and the
origin of  fermion masses,
 seems to describe
 these phenomena correctly, but it has a
naturalness problem, since too much fine tuning of the Higgs coupling
is required in order to
keep the renormalized Higgs mass acceptably small. One of the alternatives
 proposed in order to circumvent this problem is provided
by supersymmetry, a ``weak coupling'' alternative,
which introduces additional Higgs fields, but at the
same time solves the naturalness problem. The other one is technicolor,
a ``strong coupling'' alternative, which postulates the existence of new
fermions, called technifermions, which interact strongly with each other
via a
technicolor gauge interaction. In such a framework, the role of the Higgs
fields is played by condensates of
technifermion pairs. The subject of the present
paper is centered on this second
alternative.

Technicolor theory is based on the {\em ad hoc} introduction of $N_{f}$
new fermions, initially massless, not experimentally detected yet,
and having a new quantum number called
technicolor \cite{TC}.
The gauge group responsible for their mutual interactions,
 traditionally called technicolor group,
 leads to confinement and to the dynamical break-down of the initial
global chiral $SU(N_{f})_{L}
\times SU(N_{f})_{R}$ symmetry down to $SU(N_{f})_{V}$.
Due to Goldstone's theorem, this leads to the appearance of $N^{2}_{f} - 1$
massless Goldstone bosons. In such a scenario, three of them are ``eaten''
 by the electroweak gauge bosons
 $W^{\pm}, Z^{0}$,
 and the rest $N^{2}_{f} - 4$ become pseudo-Goldstone bosons (PGBs),
 after acquiring masses due to the explicit chiral
 symmetry breaking by
the conventional Standard-Model interactions
$SU(3) \times SU(2)_{L} \times U(1)$. These PGBs are composite particles,
 consisting of 2 technifermions,
and they are singlets under the technicolor group. If
this mechanism causes the breaking of the electroweak symmetry, the order of
magnitude of the scale $\Lambda_{TC}$ of
the technicolor group,
 where confinement of the  technifermions occurs, should be on
the order of 1 TeV.

Even though the above mechanism can explain the masses of the gauge bosons of
weak interactions, it does not explain the masses of ordinary fermions.
This problem is solved by postulating the existence of a new interaction,
called extended technicolor (ETC), that is associated with a gauge group
that is broken at an energy scale $\Lambda_{ETC}$, usually
 much larger than $\Lambda_{TC}$.
 Both ordinary fermions and technifermions feel this
interaction, which, at scales close to 1 TeV, manifests itself
in the form of effective (non-renormalizable)
4-fermion interactions among fermions and technifermions.
Thus, a condensate of two technifermions (T) can ``feed down'' its mass to
ordinary fermions (f), via an interaction of the form

\begin{equation}
\frac{\lambda^{2}_{ETC}}{\Lambda^{2}_{ETC}}\bar{f_{L}}T_{L}\bar{T_{R}}f_{R},
\end{equation}

\noindent where $\lambda_{ETC}$ is the effective ETC coupling.
 The fermion masses are then given by

\begin{equation}
m_{f} \approx  \frac{\lambda^{2}_{ETC}}{\Lambda^{2}_{ETC}}<\bar{T}T>.
\label{eq:ferma}
\end{equation}

However, one should also expect effective ETC interactions of the form

\begin{equation}
\frac{1}{\Lambda^{2}_{ETC}}\bar{f_{L}}f_{L}\bar{f_{R}}f_{R},
\end{equation}

\noindent which could potentially
lead to problems with too large flavor-changing
neutral currents (FCNC). In order
to avoid that, the ETC scale $\Lambda_{ETC}$ must be taken very large, on
the order of about 1000 TeV. This leads to very small fermion masses,
according to Eq.\ref{eq:ferma}, and it
certainly cannot account for the masses of the heavier quarks, unless an
excessive fine tuning of the effective ETC coupling is used. This would
unfortunately lead us back to the naturalness problem, a problem that
technicolor was created in order to avoid. A solution to this problem was
proposed some years ago \cite{walk}, in the form of ``walking" technicolor
models, in which the technicolor gauge coupling runs slowly due to
the screening of the technicolor charge by the technifermions.
This mechanism allows for large fermion masses, while adequately
suppressing FCNC.

During the last decade, the testing of the phenomenological
consequences of technicolor models made it very important
to study very carefully the momentum dependence of the
self-energy of the technifermions, i.e. the transition
from the ``constituent" technifermion masses, at low energies,
to the ``current" technifermion masses, at high energies.
A way of performing such a study is provided by the
CJT formalism \cite{CJT}, which was originally developed
in the QCD context, but can in principle be applied to any
strongly interacting theory. In this formalism, one starts
from a Lagrangian describing the strongly interacting fermions,
 and after using
effective action techniques one writes down a Schwinger-Dyson
equation for the self-energies of these fermions.
The analysis here below follows closely Ref.\cite{TC}.

For a Lagrangian density of the form

\begin{equation}
\em{ L} = \bar{\psi}\gamma_{\mu}(i\partial^{\mu} - gA^{\mu})\psi -
\frac{1}{{\rm 4}}F_{\mu\nu}F^{\mu\nu},
\end{equation}

\noindent where $F_{\mu \nu} = \partial_{\mu}A_{\nu} - \partial_{\nu}A_{\mu}
-[A_{\mu},A_{\nu}]$ is the curvature of the technicolor gauge group, and
by $\psi$ we denote the technifermion fields,
this formalism gives us the following Schwinger-Dyson equation,
 in the ladder approximation, and neglecting the
running of the gauge coupling g:

\begin{equation}
S^{-1} = p^{\mu}\gamma_{\mu} + ig^{2}C_{2}(R)\int \frac{d^{4}k}{(2\pi)^{4}}
\gamma_{\mu}S(k)\gamma_{\nu}\frac{g^{\mu\nu}-(1-\xi)
\frac{(p-k)^{\mu}(p-k)^{\nu}}{(p-k)^{2}} }{(p-k)^{2}},
\end{equation}

\noindent where S is the fermion propagator,
$\xi$ allows for different gauge choices, and
$C_{2}(R)$ is the quadratic Casimir invariant
of the fundamental representation of the technicolor group. For $SU(N_{TC})$,
$C_{2}(R) = \frac{N_{TC}^{2} - 1}{2N_{TC}}$.

By making now  the ansatz
 $S^{-1} = A(p^{2})\gamma^{\mu}p_{\mu} - \Sigma(p^{2})$,
 where $\Sigma(p^{2})$ is the fermion self-energy, and in Euclidean
 space, after angular
integration, we get a set of equations:

\begin{eqnarray}
A(p^{2}) & = & 1 + \frac{\xi a }{3}\int^{\Lambda^{2}}_{0}dk^{2}
\frac{k^{4}}{M^{4}}\frac{A(k^{2})}{A^{2}(k^{2})k^{2} + \Sigma^{2}(k^{2})}
\nonumber \\ & & \nonumber \\
\Sigma(p^{2}) & = &  \frac{(3+\xi) a }{3}\int^{\Lambda^{2}}_{0}dk^{2}
\frac{k^{2}}{M^{2}}
\frac{\Sigma(k^{2})}{A^{2}(k^{2})k^{2} + \Sigma^{2}(k^{2})},
\label{eq:SDDD}
\end{eqnarray}

\noindent where $M = {\rm max}(p,k)$, $a = \frac{\alpha}{4\alpha_{c}}$, with
$\alpha = g^{2}/4\pi$ and $\alpha_{c} = \pi/3C_{2}(R)$. We have
placed a UV cut-off to our theory. In technicolor theories,  $\Lambda$ is
the typical ETC scale.
 In the Landau gauge, where $\xi = 0$, we have $A = 1$.
 If we want to have a gauge coupling strong enough
to break dynamically the chiral symmetry of the theory, and at the same
time small enough to justify the use of perturbation theory
in the CJT formalism, the relation
$1/4 \;< \;a \; < \;1$ must hold.

 In what is called
  the ``dressed" ladder approximation, we have
  a more physical situation, by allowing the gauge coupling to run, so
that
$\alpha = \alpha((p-k)^{2})$. In a certain approximation then,
we get the same equations as above, but now the coupling $a$ appears inside the
 integrals over k.

Now, for an effective, non-renormalizable Lagrangian that
contains 4-fermi interactions, like ETC interactions, of the form

\begin{equation}
{\em L} = \bar{\psi}i\partial_{\mu}\gamma^{\mu}\psi + \frac{g}{2}
(\bar{\psi}\gamma_{\mu}\psi)(\bar{\psi}\gamma^{\mu}\psi),
\end{equation}

\noindent the CJT formalism gives an inverse fermion propagator of the form

\begin{equation}
S^{-1} = p^{\mu}\gamma_{\mu} + ig\int \frac{d^{4}k}{(2\pi)^{4}}
\gamma_{\mu}S(k)\gamma^{\mu},
\end{equation}

\noindent so we can write $S^{-1}(p^{2}) = p^{\mu}\gamma_{\mu} - \Sigma$,
 where $\Sigma$ is independent of p. We then get an equation for the
 fermion self-energy $\Sigma$:

\begin{equation}
\Sigma = \lambda\int^{\Lambda^{2}}_{0} dk^{2}\frac{k^{2}}{\Lambda^{2}}
\frac{\Sigma}{k^{2} + \Sigma^{2}},
\label{eq:4fsd}
\end{equation}

\noindent with $\lambda = \frac{g\Lambda^{2}}{4\pi^{2}}$, and $\Lambda$ is a
UV cut-off in the theory, necessary in order
to avoid the logarithmic divergence of $\Sigma$.

 Combining the results obtained above, we can study a technicolor theory
that contains both 4-fermi and gauge interactions. In the Landau
gauge and in Euclidean space,  we get

\begin{eqnarray}
\Sigma(p^{2}) & = &\int^{\Lambda^{2}}_{0}dk^{2}a(M^{2})\frac{k^{2}}{M^{2}}
\frac{\Sigma(k^{2})}{k^{2} + \Sigma^{2}(k^{2})}
+ \nonumber \\ & & \nonumber \\ & + &
\lambda_{TT}\int^{\Lambda^{2}}_{0}dk^{2}\frac{k^{2}}{M^{2}}
\frac{\Sigma(k^{2})}{k^{2} + \Sigma^{2}(k^{2})} +
\lambda_{Tf}\int^{\Lambda^{2}}_{0}dk^{2}\frac{k^{2}}{\Lambda^{2}}
\frac{m_{f}}{k^{2} + m_{f}^{2}}
\nonumber \\ & & \nonumber \\
m_{f} & = & \lambda_{ff}\int^{\Lambda^{2}}_{0}dk^{2}\frac{k^{2}}{\Lambda^{2}}
\frac{m_{f}}{k^{2} + m^{2}_{f}} +
\lambda_{fT}\int^{\Lambda^{2}}_{0}dk^{2}\frac{k^{2}}{M^{2}}
\frac{\Sigma(k^{2})}{k^{2} + \Sigma^{2}(k^{2})},
\end{eqnarray}

\noindent where $\Sigma(p^{2})$ is the the technifermion self-energy, and
$m_{f}$ is the mass of an ordinary fermion
coupled to the technifermion via 4-fermi ETC interactions.
$\Sigma(p^{2})$ receives contributions
from the technicolor
gauge interactions, as well as from ETC 4-fermi interactions of the
technifermion with
other technifermions (see term proportional to $\lambda_{TT}$), and with
ordinary fermions (see term proportional to $\lambda_{Tf}$). The mass of the
fermion receives contributions only by 4-fermi interactions
of the fermion with other
ordinary fermions (see term proportional to $\lambda_{ff}$), and with
technifermions (see term proportional to $\lambda_{fT}$).

After making the approximation

\begin{equation}
\int^{\Lambda^{2}}_{0}dk^{2}\frac{k^{2}}{\Lambda^{2}}
\frac{m_{f}}{k^{2} + m^{2}_{f}} \approx
m_{f}\left(1-\frac{m^{2}_{f}}{\Lambda^{2}}
\ln{\left(\frac{\Lambda^{2}}{m^{2}_{f}}\right)}\right)
\approx m_{f},
\end{equation}

\noindent which is valid for scales $\Lambda$ large compared to the fermion
mass,
 we get the gap equations

\begin{eqnarray}
\Sigma(p^{2}) & = & \int^{\Lambda^{2}}_{0}dk^{2}a(M^{2})\frac{k^{2}}{M^{2}}
\frac{\Sigma(k^{2})}{k^{2} + \Sigma^{2}(k^{2})}
+
\lambda_{T}\int^{\Lambda^{2}}_{0}dk^{2}\frac{k^{2}}{\Lambda^{2}}
\frac{\Sigma(k^{2})}{k^{2} + \Sigma^{2}(k^{2})}
\nonumber \\ & & \nonumber \\
m_{f} & = &
\lambda_{f}\int^{\Lambda^{2}}_{0}dk^{2}\frac{k^{2}}{\Lambda^{2}}
\frac{\Sigma(k^{2})}{k^{2} + \Sigma^{2}(k^{2})},
\label{eq:SDTC}
\end{eqnarray}

\noindent with $\lambda_{f} \equiv \frac{\lambda_{fT}}{1-\lambda_{ff}}$
and $\lambda_{T} \equiv \lambda_{TT} +
\frac{\lambda_{Tf}\lambda_{fT}}{1- \lambda_{ff}}$.
We note that the second term of the right-hand side of the equation
giving $\Sigma(p^{2})$ is momentum independent, so it
enters in the problem through the boundary conditions of the
differential equation corresponding to Eq.\ref{eq:SDTC}.
Having arrived at this result, we are ready to proceed in
 a careful analysis of the behavior of these integral equations.
In the next
section, we try to analyse analogous integral equations, which can
be viewed as simplifications of the above relations, and which permit
us to study their behavior in a simple way.

\section {The role of integral equations in
 technicolor models}
 \subsection{Analytical study}

In the following, we try to describe the problem
of dynamical mass generation in an order of growing sophistication.
The simplest way that an integral gap equation
enters in the problem of mass
generation is through the Nambu-Jona-Lasinio model
 \cite{NJL}. There, we are confronted
with an equation of the form

\begin{equation}
m = \frac{a}{\Lambda^{2}}\int_{0}^{\Lambda^{2}}
\frac{mp^{2}dp^{2}}{p^{2}+m^{2}},
\label{eq:NJLeq}
\end{equation}

\noindent where m is the mass of the fermion,
which is taken to be momentum-independent,
 $\Lambda$ is the UV cut-off of the theory, and $a$ is
a coupling associated with 4-fermion interactions. This is
essentially the same as Eq.\ref{eq:4fsd}.
 Apart from the trivial solution
$m = 0$, the above equation possesses a solution
given by the equation
\begin{equation}
\frac{1}{\lambda} = 1-\frac{m^{2}}{\Lambda^{2}}
\ln{\left(\frac{\Lambda^{2}}{m^{2}}+1\right)}.
\end{equation}

\noindent Such an equation exhibits a critical behavior,
 since, for $\lambda < 1$, it does
not possess a non-trivial solution.

If we want to account for
momentum-dependent masses, things get more complicated.
One has to resort to
the Schwinger-Dyson-equations formalism, and the resulting
equations are quite
intractable. As we saw in the previous section, one
 usually has to make the ladder approximation
and go to Euclidean space and to the
Landau gauge in order to simplify the equations, which are
still analytically
solvable only in the high- and low-energy limits.

Even though
all these
manipulations make us view the final results with scepticism, it is possible
that they describe the qualitative behavior of the
theory correctly. However,
 since their precise form is questionable, we will try to analyse
  a  different, but very similar, integral gap equation,  that
  could be relevant to our problem
\footnote{ It is worth noting that a very similar analysis can be done by
 using the differential form of the gap equations, which yields the same
 results.}. Namely, we are going to follow the reverse procedure
 from the usual one, by
 making a particular ansatz for the functional form of the fermion
 self-energy,
 because we feel that it sheds light on some other
 aspects of dynamical mass generation. In that way,
we will be able to concentrate on the quantization condition, instead of being
lost in complicated gap equations, which are of questionable validity anyway.

First, we make an ansatz for the momentum-dependent fermion self-energy,
$\Sigma(p^{2})$, by assuming that, for $p \gg \Sigma_{0}$, where
 $\Sigma_{0}$ is the value of $\Sigma(p^{2})$ near the chiral symmetry
 breaking scale, it takes the form
  $\Sigma(p^{2}) \approx \Sigma_{0}(p^{2}/\Sigma_{0}^{2})^{-\gamma}$.
   We can then try
   to find what kind of integral equation a function like
$\Sigma(p^{2})$ satisfies, and then try to motivate it physically.

We first give two useful identities
\begin{eqnarray}
\int^{\infty}_{0}e^{-st}t^{\gamma-1}dt & = &
\Gamma(\gamma)s^{-\gamma} \\ \nonumber && \\ \nonumber
\int^{\infty}_{0}e^{-st}t^{-\gamma}dt & = &
\Gamma(1-\gamma)s^{\gamma-1},
\end{eqnarray}

\noindent where $\Gamma$ is the usual $\Gamma$-function, and
$0 < {\rm Re}{(\gamma)} < 1$. Combining the above equations, we get
 an eigenvalue integral equation of the form
\begin{equation}
G(s) = \tilde{\lambda}\int^{\infty}_{0}e^{-st}G(t)dt,
\label{eq:simple}
\end{equation}

\noindent with eigenfunction $G(s)  = \left(\sqrt{\Gamma(\gamma)}
s^{-\gamma} - \sqrt{\Gamma(1-\gamma)}s^{\gamma-1}\right)$,
 and eigenvalue
\[ {\tilde \lambda} =
\left(\sqrt{\Gamma(\gamma)\Gamma(1-\gamma)}\right)^{-1} =
\sqrt{\frac{\sin{(\pi\gamma})}{\pi}}.\]
So, even though the functions $s^{-\gamma}$ and $s^{\gamma - 1}$ are not
separately eigenfunctions of the integral operator
 $\int_{0}^{\infty}e^{-st}[...]dt$, their specific linear
combination given above is an eigenfunction.
Unfortunately, it is not clear how one
could interpret physically the integration measure
$e^{-st}dt$ in Eq.\ref{eq:simple}.

By applying the same integral operator twice,
 however, we end-up with the following equation:
\begin{equation}
G(s) = \lambda \int^{\infty}_{0}\frac{G(t)}{s+t}dt,
\label{eq:central}
\end{equation}

\noindent where $G(t)$ is the same as above,
again with $0 < {\rm Re}{(\gamma)} < 1$,  and $\lambda =
\frac{\sin{(\pi\gamma)}}{\pi}$.
In purely mathematical terms, we are dealing now
with a function $G(s)$ which, up to a numerical
coefficient, which is the
eigenvalue of the equation,
is its own simple Stieltjes transform.
This time, the
integration measure  $\frac{dt}{s+t}$ is much
 easier to interpret physically.
 Making the correspondence of $G(s)$ with the fermion
 self-energy, we can also make the correspondence of
 the s and t variables with  squares of
4-momenta, say $s=p^{2}$ and $t =k^{2}$. Then,
this integration measure is
very close to the one appearing in the Schwinger-Dyson
(S-D) equations in Eq.\ref{eq:SDTC}.

More
precisely, the integration measure before
performing the  angular integration  that gives
Eq.\ref{eq:SDTC} is of the form (see, for instance, Ref.\cite{TC})
$\frac{1}{\pi^{2}(p-k)^{2}}\frac{dk^{4}}{k^{2} + \Sigma^{2}(k^{2})}$.
 After angular integration, where spherical symmetry of the self-energy is
 assumed, the term $(p-k)^{2}$ is replaced by the
  quantity $M \equiv max(p^{2}, k^{2})$ (see Eq.\ref{eq:SDTC})
  \footnote{Even though this replacement is exact in the
  case of a non-running gauge coupling, it is just an approximation in the
  running case.} . The
integration measure appearing in Eq.\ref{eq:central} is
 equivalent to
approximating M with the quantity
$p^{2}+k^{2}$, and, in addition,
neglecting the self-energy $\Sigma(k^{2})$
appearing in the denominator. Therefore,
Eq.\ref{eq:central} is a linearized version of the usual
 S-D equations, and it is expected to give trustworthy results only in
 the limit $p^{2} \gg \Sigma^{2}_{0}$.
 In that limit, our integration measure
 is a special case of a more general
 form that has appeared in the literature, giving similar results
 \cite{kerbob}.
In such a context, one should not worry about  the small-s  behavior
of the eigenfunction G(s), which possesses a singularity in that region;
a singularity which is integrable but nevertheless unphysical.

In all that discussion, we also neglect the extended technicolor contributions
to the technifermion self-energy, since
their effect can be usually absorbed in the UV boundary conditions.

An interesting feature of Eq. \ref{eq:central} is that,
in general, the two terms $t^{-\gamma}$ and $t^{\gamma -1}$
 are separately eigenfunctions of the same integral operator
 and the same eigenvalue. It is not necessary to take the
 specific linear combination used in Eq.\ref{eq:simple}
 any more. However, if we take the exponent $\gamma$ to
  be complex, a situation that will appear in the next
  sections of this paper, and we insist on having a real
   eigenfunction, the situation changes:
  we then  require  $\gamma$ to be of the form
  $\gamma = 1/2 + i\delta$, with $\delta$ a real
   number. In addition,
   we have to
   keep the linear combination that we have used so far,
   since $(\Gamma(\gamma))^{*} = \Gamma(1-\gamma)$, or,
   for this matter, any linear combination of the two
   terms, where one of the coefficients is the complex conjugate
of the other.

It is also interesting to note that  the form of the eigenfunction
$G(s)$ given above is not the most general one corresponding
to the Stieltjes kernel in the special case where $\gamma = 1/2$. In
that case, a more general form is $c_{1}s^{-1/2}+c_{2}s^{-1/2}\ln{(s)}$,
where $c_{1,2}$ are arbitrary constants. The eigenvalue associated with
this solution is exactly the same as the one for general $\gamma$.
We are not going to occupy ourselves with this special case any further.

We can try to recover now  the form of the solutions that the
S-D equations give in Ref.\cite{TC}. In that analysis, the coupling $\lambda$
satisfies the relation
 $\lambda = \frac{\alpha}{4\alpha_{c}}$, where
 the couplings are taken to be momentum independent for the moment.
  That means that, since in our case the coupling and the exponent
  $\gamma$ of the eigenfunction are related by the equation
$\lambda = \sin{(\pi\gamma)}/\pi =
\cosh{(\delta\pi)}/\pi$, in the regime
$\delta\pi \ll 1$  we get $\delta \approx
\frac{2}{\pi}\sqrt{\lambda\pi - 1} =
\frac{2}{\pi}\sqrt{\frac{\pi\alpha}{4\alpha_{c}}-1}$.
 The quantity corresponding to $\delta$ in the exponent of
 the eigenfunctions used in Ref.\cite{TC}
  is equal to
$\sqrt{\alpha/\alpha_{c}-1}/2$, which is reasonably close
 to our expression, given the different integration measure
  used in the two cases, and the fact that we expanded our
   expression for the coupling for
small $\delta$, i.e. for $\lambda \approx 1/\pi$.
Moreover, for the proposed form of the quantity $\gamma$,
the above eigenvalue equation exhibits a critical behavior
\footnote{ Note that $\cosh(x) > 1$, for all real numbers x.},
since, for real values of $\delta$,  it
possesses non-trivial solutions only for
$\lambda > 1/\pi \approx 0.32$.

Finally,
by taking the factor multiplying one of the two solutions
to be equal to $\Sigma_{0}^{2}e^{i2\delta\theta}$,  and taking
care of the correct dimensionality of the quantities used,
the resulting expression for the self-energy is

\begin{equation}
\Sigma(p^{2}) = \frac{\Sigma_{0}^{2}}{p}
\sin\left[2\delta(\ln(p/\Sigma_{0})+\theta)\right],
\label{eq:self}
\end{equation}

\noindent
a well-known functional form in the technicolor
literature \cite{TC}. Here $\Sigma_{0}$ is the characteristic
energy of the theory, which is on the order of the chiral
symmetry breaking scale.

Unfortunately, such a solution in general
possesses nodes, and momentum regions
at which the self-energy becomes negative. It is unclear how one could
interpret such solutions physically. We will encounter particular
examples of the behavior of such a solution later in the paper.

\subsection{The running coupling case}

In technicolor theories, one typically has a technicolor
 non-abelian group, with a coupling $\alpha(p^{2})$ that
  is renormalized, and an extended technicolor group
  that is broken at very high energies (on the order
  of the ETC scale), with an ETC coupling that is taken to be constant.
We assume that the ETC coupling is below the critical
 value that would allow it to break chiral symmetry without
  the need of technicolor gauge interactions.
  Then, chiral symmetry will be broken
at a scale where the running  coupling $\alpha(p^{2})$
 gets strong enough, so that it, together with the ETC-coupling,
  can bring the system to criticality. For energies above that scale though,
 the gauge coupling is below its critical value. In that case,
 we expect the form of the solutions of the Schwinger-Dyson
  equations to change.
  First of all, we should insert the coupling $\lambda$ inside
  the integral sign. Then, we can expect that, in a crude
  approximation, the relation between the coupling and the
  exponent of the function $G(t)$ will remain the same as for
  the constant coupling case.

 In
other words, we expect the quantity $\delta$ defined above
to become purely imaginary, so that  the form of the
 solution for the fermion self-energy  becomes
\begin{equation}
\Sigma(p^{2}) = \frac{\Sigma^{2}_{0}}{p} (p/\Sigma_{0})^{2\tilde{\delta}},
\label{eq:self2}
\end{equation}

\noindent where $\tilde{\delta} = i\delta =
\sqrt{1- \frac{\pi\alpha(p^{2})}{4\alpha_{c}}}$.

 Such a naive analysis, however, neglects the complications
 arising from the fact
 that the coupling is running. There have been more careful analyses
 of the form of the self-energy and the way it evolves up to the
 ETC scale
 (see, for instance, \cite{TC}), and they have shown that,
 in general, the self energy at the ETC scale has approximately
 the form

 \begin{equation}
 \Sigma(\Lambda_{ETC}^{2}) \approx
 \frac{\Sigma_{0}^{\omega}}{\Lambda_{ETC}^{\omega-1}},
 \label{eq:enhance}
 \end{equation}

\noindent where  the power $\omega$ can be anywhere between 3 - the
 case of running coupling - and 2 - the limiting case of walking coupling,
  where we have an non-trivial UV fixed point in the theory.
 The parameter $\omega$ can also approach 1 in theories where
 the high-momentum enhancement is coming from 4-fermion interactions \
 \cite{TC}, which, as we have already seen,
 influence the boundary conditions of
 Eq.\ref{eq:SDTC}.

 The parameter $\omega$ is related in a very complicated way to $\delta$,
  or, in other words, to
  the coupling $\lambda$. The reason for this complexity is the running
 of the gauge coupling, and the introduction of ETC 4-fermi couplings can
 make the situation even worse. The thing we can say here for sure is that
 we expect $\omega$ and $\delta$ to be negatively correlated, i.e. a larger
 coupling, over a large momentum region, should correspond to a smaller
 $\omega$.
 This is intuitively reasonable, since
 usually $\Sigma_{0}$ is much smaller than $\Lambda_{ETC}$, and
  a larger coupling at large momenta should be able to
 produce larger self-energies $\Sigma_{0}(\Lambda^{2}_{ETC})$. When studying
  the full
 non-linear Schwinger-Dyson equation, we also expect  the parameter $\theta$
 to be a function of $\delta$, which is nevertheless too complicated to
 be computed analytically.
 The relation in Eq.\ref{eq:enhance} is going to
 be  frequently used in the next section.

\subsection { The quantization condition}
The previous subsection dealt extensively with Eq.\ref{eq:central}
and its close connection to Eq.\ref{eq:SDTC}. Eq.\ref{eq:central}
 is a homogeneous Fredholm equation of the
second kind. Unfortunately,
as it stands, its kernel does not belong to $L_{2}$, since the double integral
\begin{equation}
\int_{0}^{\infty}dt\int_{0}^{\infty}ds\; \frac{1}{(s+t)^{2}}
\end{equation}
diverges. Therefore, it is not possible to apply the usual Fredholm theorems
in this case \cite{Tricomi}.
An example of the singular behavior of
Eq.\ref{eq:central} is that, as we saw, the eigenvalues
 associated with it belong to a continuous spectrum of eigenvalues.
 The divergence of this double integral is logarithmic,
 and it comes from both the ultra-violet (UV) and
 infra-red (IR) regions. In both these cases, however, there are
 physical cut-offs that render the kernel square-integrable.

 First of all, there is a UV-cut-off $\Lambda$ associated with the new physics
 coming in at that scale.
  In technicolor models, for instance, the
 infinite upper bound of integrations of this kind is
  usually replaced by a finite cut-off $\Lambda_{ETC}$,
   where new physics in the form  of extended-technicolor
    interactions come into play.
    Moreover, the role of the IR cut-off is played by the fermion
    self-energy $\Sigma(t)$ that should appear in the denominator of our
    kernel if we had not linearized our integral equation, i.e. if
    the kernel were of the more
    physical form $\frac{t}{(s+t)(t+\Sigma^{2}(t))}$. The linearization
    of our equation, as we shall see,
     while simplifying our analysis considerably, is not going to
    affect our final results in a qualitative way.

    From the moment the
    kernel belongs to $L_{2}$, we
     should expect
     Fredholm's theorems to apply, and
     the spectrum of the eigenvalues of Eq.\ref{eq:central}
     to become discrete.
    Let's see how this mechanism works in our case.

Taking care of the correct dimensionality of the
 quantities used, we can rewrite Eq. \ref{eq:central} as
\begin{eqnarray}
c\left(\frac{s}{\Sigma^{2}_{0}}\right)^{\gamma-1} -
c^{*}\left(\frac{s}{\Sigma^{2}_{0}}\right)^{-\gamma} &=  & \lambda
\int^{\infty}_{0}dt
\left[\frac{c\left(\frac{t}{\Sigma^{2}_{0}}\right)^{\gamma-1} }{t+s}
- \frac{c^{*}\left(\frac{t}{\Sigma^{2}_{0}}\right)^{-\gamma}}{t+s}\right] =
\\ \nonumber && \\ \nonumber
 &&= \lambda\int^{\Lambda_{ETC}}_{0}dt
\left[\frac{c\left(\frac{t}{\Sigma^{2}_{0}}\right)^{\gamma-1} }{t+s}
- \frac{c^{*}
\left(\frac{t}{\Sigma^{2}_{0}}\right)^{-\gamma}}{t+s}\right] + I(s),
\end{eqnarray}

\noindent where $I(s)= \lambda\int_{\Lambda_{ETC}}^{\infty}dt
\left[\frac{c\left(\frac{t}{\Sigma^{2}_{0}}\right)^{\gamma-1} }{t+s}
- \frac{c^{*}\left(\frac{t}{\Sigma^{2}_{0}}\right)^{-\gamma}}{t+s}\right] $,
and $\Sigma_{0}$ is the typical energy scale of the model, i.e. the
value of the fermion self-energy at low energies, which is on the order
of magnitude of the chiral symmetry breaking scale.
We are now going to impose the condition $I(s)= 0$, and
we are going to investigate what constrains
such a condition imposes on the solutions.

By
performing the above integrals, we have

\begin{eqnarray}
&&I(s)= 0 \Longrightarrow  \\ \nonumber  \Longrightarrow &
&\frac{c\left(\frac{\Lambda_{ETC}}{\Sigma_{0}}\right)^{2(\gamma-1)}}
{1-\gamma} F(1, 1-\gamma;2-\gamma;-s/\Lambda^{2}_{ETC}) +
\\ \nonumber && \\ \nonumber &+&
\frac{c^{*}\left(\frac{\Lambda_{ETC}}{\Sigma_{0}}\right)^{-2\gamma}}
{\gamma} F(1, \gamma;1+\gamma;-s/\Lambda^{2}_{ETC})  = 0,
\label{eq:hyper}
\end{eqnarray}

\noindent where F(a, b; c; z) is the usual hypergeometric
function.
Note that, had we included a momentum-independent self-energy term
$\Sigma^{2}_{0}$
 in the
denominator of our kernel, in order to formally maintain the kernel in
$L_{2}$, the only change would be in the argument z of the hypergeometric
function, which would go from $-s/\Lambda^{2}_{ETC}$ to
$-(s+\Sigma^{2}_{0})/\Lambda^{2}_{ETC}$.
It is therefore seen that the inclusion of
such a term cannot alter our results substantially
in the region of interest, which is $s \gg \Sigma^{2}_{0}$, and this is
 expected to remain true
even if we insert in the denominator
a more realistic momentum-dependent self-energy
$\Sigma(t)$ \footnote{This would lead us to the full, non-linear equation,
where any discussion on eigenvalues and their spectrum is meaningless.
We may apply our analysis, however, to the case of large momenta.}.

The quantization condition that should derive from our equations
should be of course momentum independent. In order to simplify
our problem, we are going to make two
different approximations that will enable us to derive such a condition
in two different  momentum regimes.
 First, we restrict the momentum regime
  to $\Sigma^{2}_{0} \ll s \ll \Lambda^{2}_{ETC}$,
which of course implies also that
$\Sigma^{2}_{0} \ll \Lambda^{2}_{ETC}$. Then,
we can keep only the zeroth-order term of the series
 expansion of the hypergeometric function, and,
 setting $c = \| c \| e^{i2\delta\theta}$,
with $0 < 2\delta\theta < 2\pi $, we have the
momentum-independent equation
\begin{equation}
\frac{e^{i2\delta\theta}}{1/2 - i\delta}
\left(\frac{\Lambda_{ETC}}{\Sigma_{0}}\right)^{i2\delta} +
\frac{e^{-i2\delta\theta}}{1/2 + i\delta}
\left(\frac{\Lambda_{ETC}}{\Sigma_{0}}\right)^{-i2\delta} = 0,
\end{equation}

\noindent which is equivalent to the relation
\begin{equation}
2\delta\theta + \arctan{(2\delta)} +
2\delta\ln(\Lambda_{ETC}/\Sigma_{0}) = n\pi,
\label{eq:quantization}
\end{equation}

\noindent where n is an integer.

This quantization condition has been
derived previously, using different techniques
 (see, for instance, Ref.\cite{TC}). If now, instead of taking the limit
  $s \ll \Lambda^{2}_{ETC}$, we take the limit
   $s \rightarrow \Lambda^{2}_{ETC}$, we get a similar
   quantization condition, but with the term $\arctan{(2\delta)}$
   replaced by the constant $\pi/2$. This difference
   is considered to be an artifact of
   our derivation and the approximations involved in it.
   Moreover, it is not a significant change, and it is
   not expected to alter the qualitative aspects of our results
   below, which rely mostly on the term $\ln(\Lambda_{ETC}/\Sigma_{0})$.
   Since our quantization condition is going to be mostly used at energies
   $s \approx \Lambda^{2}_{ETC}$, we are going to use the form with the
   $\pi/2$ factor in it.

   One could object that these results are not reliable, since
in $I(s)$ we assume that the self-energies retain the same functional forms
 for $p > \Lambda_{ETC}$
 as for momenta below $\Lambda_{ETC}$. However, a similar, more physical
but more complicated
analysis, using Heavyside functions which truncate the eigenfunctions at
momenta above the cut-off, yields exactly the same results. Such an analysis
was used recently on a completely different physical context \cite{Lands},
 in order to derive a discrete eigenvalue spectrum out of an integral
 equation.

In the full non-linear theory, $\theta$ is in principle a function
of $\delta$. In our linearized equations, however, $\theta$ can be
arbitrary.
Since we have assumed that
$\Sigma^{2}_{0} \ll \Lambda^{2}_{ETC}$ and that $\theta > 0$,
from Eq.\ref{eq:quantization} we see that
n has
 to be a positive integer larger than or equal to 1, i.e. $n \geq 1$.
The relation appearing in Eq.\ref{eq:quantization} provides us with
a quantization condition, which is going to be central to the
development of this paper. It has been previously derived using other methods,
but solutions for $n > 1$ have not been really exploited.

Since, in the most general case,
we are dealing with a non-abelian gauge group with a gauge coupling that
is renormalized, the use of a constant gauge coupling as above
is not very realistic.
Therefore, in the discussion that follows,
 we are going to study
 the running coupling case.

As soon as we have to cope with a running gauge coupling, however, the
situation changes dramatically. For the parameter $\delta$ is real
at low momenta, but as we go to larger momenta it becomes imaginary.
The problem is that the quantization condition that we derived previously
is based on the assumption that $\delta$ is real (and constant).
In addition, in the running
coupling case, $\delta$ is real in the momentum region where
non-linearities become important, and the notion of the eigenvalue spectrum
becomes problematic.
The equation becomes so
difficult for running $\delta$ that we were not able to
compute analytically a quantization
condition. Nevertheless, we know that, since
the kernel of the linearized equation still
 belongs to $L_{2}$ \footnote{we assume here that the gauge coupling
 does not possess a non-integrable
 singularity, but stays at finite values instead.},
a quantization condition must exist, since the spectrum of eigenvalues
 must be
discrete. To make a very crude approximation, we are going to assume that
the same quantization condition as above
is also valid  for the running case, where
in the place of $\delta$,  a real
parameter $\delta$ is used
in some ``average" sense.

 The use of such a parameter is based on the argument that,
 since there is
 chiral symmetry breaking and our equation possesses non-trivial
 solutions even in the running-coupling case, in a certain
 ``average" sense we may consider  $\delta$
  to be
 real, i.e. the gauge coupling is above its critical value.
 This does not stop the actual
 $\delta$ parameter
 to reach imaginary values at large momenta.
 Actually, in the integral of Eq.\ref{eq:hyper}, the actual
 $\delta$ parameter is imaginary throughout the whole integration
 region. However, as we said previously, there are alternative
 ways for deriving exactly the same quantization condition, while
 staying inside the physical momentum region $0 \leq p \leq \Lambda_{ETC}$.
 The crudeness of
 such an analysis should
 not obscure the fact that the very nature of our equations,
  in the high-momentum, linear regime,
 makes them obey
 a certain quantization condition, even though it proves non-trivial,
 if not impossible,
 to find its exact analytical form. For instance, it is conceivable that, in
 the running coupling case, the ratio $\Lambda_{ETC}/\Sigma_{0}$ has a
 dependence on the integer n
 that is closer to a power law, instead of an exponential law, as implied
 by Eq.\ref{eq:quantization}. We do not try to analyse this, or other similar
 possibilities, here, as it would further complicate our analysis.

In the next section, we are going to
analyse carefully the effects of the quantization condition,
manipulated in the way described above, on the
self-energy $\Sigma(p^{2})$.

\section { Physical interpretation}
In this section, an attempt is made to find what physical
consequences Eq.\ref{eq:quantization} can have.
In particular, we would like to see if the above quantization
condition is in any way  related to the appearance of
 the known fermions in three different generations.
  We cannot help remarking that the boundedness of operators
     similar to the one studied here  is the source
     of quantization in ordinary
quantum mechanics, like the energy
levels of an electron confined in a finite box, or the energy levels of the
hydrogen atom.
  In order to see
if a mechanism of this sort could be qualitatively
realistic in our case, we are going to neglect isospin mass splitting
within the $SU(2)_{L}$ doublets, assuming that another
mechanism is responsible for it,
and we are going to consider only the up, charm and top
for the quarks, and the electron, muon and tau for the leptons.

 A very crude, order of
 magnitude inspection of their current masses
reveals a hierarchy of a factor of about 200 among each
subsequent generation, since, for the quarks,
$m_{u} \approx 5$ MeV, $m_{c} \approx 1.5$ GeV, and
  we expect the top to have
 a mass of about $m_{t} \approx 170$ GeV.
The top mass
 seems to be smaller than $200m_{c}$, but
 we should not forget that we are
 making an order-of-magnitude, qualitative
 discussion.
  This picture would correspond to linear trajectories on what
 is sometimes called the ``Bjorken plot".
 One could argue that, since
 the mass difference of the top and bottom quarks is so large,
 it is quite arbitrary to chose the upper partners of the quark doublets,
 and one could just as well
 consider the lower partners of the doublets instead, which would lead us to
 very different results. However,
 even though we do not have any rigourous argument towards that,
 we feel that, in a theory that contains a minimum number
  of adjustable parameters, the top quark is the one that
 has the most ``natural" mass, being the closest to the weak scale, where
 we believe that the fermion-mass origins lay.
 This leads us then to compare the top quark mass with the other
 two quarks having charge +2/3.

 For the leptons, we have something similar happening, since
 $m_{e} \approx 0.5$ MeV, $m_{\mu} \approx 0.1$ GeV, and
 $m_{\tau} \approx 1.8$ GeV.
 We assume that the mechanism that makes $m_{\tau}$
 considerably smaller than $200m_{\mu}$ is similar to the one making the
 expected value for $m_{t}$ smaller than $200m_{c}$.
 Here, we neglect the upper partners of the
 lepton doublets, the neutrinos, leaving again to another mechanism the
explanation for
 the smallness, or the vanishing, of their masses.
  In the following, we are going to assume
 that QCD or other effects can account for the quark-lepton mass difference,
  since the proposed mechanism cannot account for it.

 Having this in
  mind for the ordinary fermions, it will be also useful
   to remind the reader that in technicolor theories, it is generally
   expected that $\Sigma_{0}$, the maximum self-energy of the
technifermions,  is on the order of the chiral symmetry
breaking scale $\Lambda_{\chi}
 \approx 1$ TeV.
  Moreover, since one expects
  the technifermions and the ordinary fermions to lie
   in the same  representation of the
extended technicolor group,  before this breaks at
 scale $\Lambda_{ETC}$,
the value of the technifermion  self-energy at the
extended technicolor scale $\Lambda_{ETC}$  is expected to be
 on the order of the current mass of the corresponding
 ordinary fermion, i.e.
$\Sigma_{Tf}(\Lambda^{2}_{ETC}) \approx m_{f}$. The
 above physical constrains are going to facilitate
 considerably the analysis of the quantization condition
 appearing in the previous section, and its possible connection to
 the fermion-generation puzzle.

Such a connection is inspired from the fact that the linear trajectories in
the ``Bjorken plot" could be attributed to some exponential
dependence of the ratio of
the two
fundamental scales in the theory,
$\Lambda_{ETC}/\Sigma_{0}$, on a quantization
integer index, as in Eq.\ref{eq:quantization}.
The fact that the top mass seems
to be smaller than what expected for linear
trajectories could be an indication
that the dependence of the ratio $\Lambda_{ETC}/\Sigma_{0}$
on the quantization index does not follow
 an exponential law, but a power law or something similar instead,
 because of a possible modification of Eq.\ref{eq:quantization}
 due to the fact that the gauge coupling is not constant. In the following,
 we do not try to modify the quantization condition, but describe an analysis
 that could be based, in principle, on other similar conditions.

 It is
 essential to notice that the fact
 that the dependence of $\Lambda_{ETC}/\Sigma_{0}$ on the coupling $\delta$,
as in Eq.\ref{eq:quantization}, is non-analytic is a consequence of the
non-perturbative nature of the Schwinger-Dyson approach that
we chose to follow. Therefore, the  results of any analysis
based on this equation cannot be replicated by any perturbative considerations.
Moreover, we should stress the fact that in what follows, we are going to
refer to the ETC scales $\Lambda_{ETC}$ in a very broad sense, and they should
rather be viewed as new physical thresholds, since the discussion is not
within the framework of conventional extended technicolor scenarios.

In Eq.\ref{eq:quantization}, three main physical parameters are
involved: the extended
technicolor scale $\Lambda_{ETC}$, the value of the
fermion self-energy at zero momentum $\Sigma_{0}$,
 and $\delta$, which is related to the coupling $\lambda$.
Another  parameter which in the full, non-linear theory
is  a function of  $\delta$,
the phase $\theta$, is also entering the picture, and its value
might have interesting consequences, as we will see later.
Therefore, it seems as if  our theory contains only
two fundamental parameters, since the third can be determined by means
of the quantization equation. One can further note that
$\delta$
is mainly determined by the scale at which the
former becomes strong, i.e. the confinement scale,
 and by the type of the non-abelian gauge group and
  the technifermion content of the theory. Therefore, if
 one  assumes that
the technicolor confinement scale is directly related to
$\Sigma_{0}$ and the chiral symmetry breaking scale, or, in other words,
the weak scale,
one is essentially left with a
single parameter, along with a choice
of the technicolor group and technifermion content, which
renders this picture quite elegant. One should not forget, nevertheless, that
our mechanism requires an additional parameter, which is the ETC effective
4-fermion coupling,
 which we take to be the same for all the fermions, and which
 influences, along with $\delta$, the value of
the power $\omega$ in Eq.\ref{eq:enhance}.

In the following, we consider $\delta$ as a free parameter that is not
related to $\Lambda_{TC}$, since in our formalism $\delta$ is used
in an ``average" sense, and the connection between the two parameters
seems non-trivial.
{\em A priori}, we can fix a value for any two of these three
 parameters, and find a discrete set of values for the third
 one.
  Let us investigate all
possible combinations.
First, we can fix a value for $\delta$ and
$\Lambda_{ETC}$, and find a
discrete spectrum for $\Sigma_{0}$.
In this case, if we still want to pursue the argument of having to deal
with essentially only one fundamental parameter in the theory, we have
to assume that it is only the first member of the $\Sigma_{0}^{(n)}$
spectrum that is directly, and in a non-trivial way, associated with $\delta$.
 $\Sigma^{(n)}_{0}$ is given by the expression

\begin{equation}
\Sigma^{(n)}_{0} = \Lambda_{ETC}
e^{\theta + \frac{\pi}{4\delta}}e^{-n\pi/2\delta}.
\label{eq:ssquant}
\end{equation}

\noindent  We then assume that
some of the solutions of this equation are related to the
self-energies, at small momenta, of the
 technifermions corresponding to the three different generations
 of ordinary fermions. Then, inserting the above relation
 into Eq.\ref{eq:enhance},  we find

 \begin{equation}
 \Sigma^{(n)}(\Lambda_{ETC}^{2}) =
  \Lambda_{ETC}e^{\omega(\theta + \frac{\pi}{4\delta})}
  e^{-\frac{n\pi\omega}{2\delta}}.
  \label{eq:squant}
 \end{equation}

Since we do not find any physical reason not to take consecutive
solutions of the above equation,
we consider the first three of them,  for $n = 1, 2, 3$, and we take
them to
correspond to the technifermions associated with the top, charm, and up
 quarks respectively.

 First of all, that would mean that we have to
choose $\delta$ in such a way that
$e^{-\frac{\pi\omega}{2\delta}} \approx 1/200$,
which is the approximate hierarchy between consecutive fermion
generations of characteristic mass scale $m_{f}^{(n)} \approx
\Sigma^{(n)}(\Lambda^{2}_{ETC})$\footnote{ By n we index the
fermion generations, and the spectrum of solutions deduced by
the quantization condition.}.
 This would imply,
with the use of Eq.\ref{eq:squant} and with a choice of a negligibly
small phase $\theta$, that
the mass of the top quark is equal to
$m_{t} \approx \Sigma^{(1)}(\Lambda^{2}_{ETC}) \approx \Lambda_{ETC}/14$.
For $m_{t} \approx 170$ GeV, this gives an ETC scale $\Lambda_{ETC}
\approx 2.4$ TeV. If we choose the values $\omega = 2$ and $\theta=0$,
 this implies, from
Eq.\ref{eq:ssquant}, that $\Sigma^{(1)}_{0}
\approx 640$ GeV, and $\delta \approx 0.59$.
The choice of this value for $\omega$ has nothing in  particular and is purely
indicative. Ideally, one should be able to derive $\omega$ from $\delta$
and from the common ETC coupling of the fermions and technifermions.

Unfortunately, there are numerous phenomenological and theoretical
 problems with such a picture.
First of all, it is not clear why we do not
observe in nature lighter fermion generations associated with the
solutions of the above equations for $n  > 3$. Moreover, the
solution for $n = 2 \;{\rm or}\; 3$ implies the existence of
 technifermions
having  small self-energies at low momenta, which should
 make them observable in present experiments.
However, we have not observed signs of their existence.
 This is a serious
 phenomenological draw-back of the mechanism described above.
Furthermore,
in this picture all the fermions are associated with the same
ETC scale. A scale of about $2$ TeV is unfortunately too low to adequately
suppress flavor changing neutral currents in the light quark sector.

Another difficulty associated with this interpretation is the
stability of such solutions. From effective-potential considerations
 (see Ref.\cite{TC}, for example), it is clear that the effective
  potential is minimized for the maximum value of the self-energy. This
  makes stable only the solution for $n= 1$, and the solutions corresponding
   to higher n are unstable. This was the original reason for
   discarding solutions corresponding to higher n.
We are now going
to continue our discussion with some other possibilities that do not
seem to possess these naturalness problems.

The next possibility we can think of is to
 fix the value of $\Sigma_{0} $ and
$\delta$, and find a quantization condition for
 the extended-technicolor scales. The relation resulting from that is
\begin{equation}
\Lambda^{(n)}_{ETC} =  \Sigma_{0}e^{-\theta-\frac{\pi}{4\delta}}
e^{n\pi/2\delta}.
\label{eq:llquant}
\end{equation}

\noindent
Inserting this expression into Eq.\ref{eq:enhance}, we get
\begin{equation}
\Sigma^{(n)}(\Lambda^{2}_{ETC}) = \Sigma_{0}
e^{(\omega - 1)(\theta+\frac{\pi}{4\delta})}
e^{-\frac{n\pi(\omega - 1)}{2\delta}}.
\label{eq:lquant}
\end{equation}

\noindent
If we want to reproduce the fermion hierarchy observed in nature,
we must require that
$e^{-\frac{\pi(\omega - 1)}{2\delta}} \approx 1/200$.
 From Eq.\ref{eq:lquant}, setting $\omega=2$ and $\theta=0$,
this would mean that $m_{t} \approx \Sigma_{0}/14$,
 so, for $m_{t} \approx 170$ GeV,
this gives $\Sigma_{0} \approx 2.4$ TeV.

In such a scenario, however,
we have to be careful not to produce an unwanted hierarchy
between the weak scale (or $\Sigma_{0}$) and the top quark
mass (or $\Sigma^{(1)}(\Lambda^{2}_{ETC})$),
since our goal is to explain the maximum number
of physical scales, using
a minimum number of input parameters and mass hierarchies.
 In order to do that, we will
have to use the phase $\theta$, which up to now has  not been
really exploited.
Taking $\theta$ to be close to $\pi/2$, and fixing $\omega = 2$,
we find $\Sigma_{0} \approx 500$ GeV.
 Quite interestingly, such a choice almost eliminates another hierarchy,
 the one
between $\Sigma_{0}$ and $\Lambda_{ETC}^{(1)}$, suggesting that the ETC scale
associated with the top quark is actually very close to  1 TeV. According
to Eq.\ref{eq:llquant}, the ETC
scales are then $\Lambda^{(1)}_{ETC} \approx 1.4$ TeV,
$\Lambda^{(2)}_{ETC} \approx 290$ TeV, and
$\Lambda^{(3)}_{ETC} \approx 58 \times 10^{3}$ TeV.
Furthermore, we find $\delta \approx 0.3$. The  ETC scale associated with
the lightest generation,  $\Lambda^{(3)}_{ETC}$, is much larger than the
ones usually used in the literature, but we do not find any physical reason
that would prevent it from getting such a high value.

Moreover, we should caution the reader one more time that our results
are purely indicative. Namely, we showed how a non-zero value of
$\theta$ could fix various scales at  reasonable values, but we should keep
in mind that, in the full, non-linear equation, $\theta$ is determined
by $\delta$, and therefore it is not a free adjustable parameter.
We may add as a speculation, that
this dependence, which in a more careful study can
be determined by the numerical solution of the integral equation, could
be responsible for the fact that $m_{t}$ is expected to
be less than about $200m_{c}$.
It should be noted that there are numerical indications
that $\theta$ is non-zero \cite{TC}.
This is an alternative way of
getting around the problem of  the deviation of the top quark mass
from the linear ``Bjorken trajectories", other than the one in which
we modify the form
of the quantization condition.

 A remark along the same lines can be written about the power
$\omega$; the fact that we took
$\omega=2$ and $\delta \approx 0.59$ in our previous example,
should have made us choose a larger value
for $\omega$  in this example, instead of using
$\omega = 2$ again, since in the present example
$\delta$ is smaller, i.e. $\delta \approx 0.3$.
A larger $\omega$ would also bring $\Lambda^{(3)}_{ETC}$ down to a smaller,
more acceptable value.
However, since we do not
know the exact dependence of $\omega$ on $\delta$, we do not feel that we
should further complicate the picture with changes in parameters that do not
add anything crucial to the qualitative behavior of the mechanism.

 This second way of looking at the quantization condition
does not seem to have the stability problem that the previous
solutions had, nor does it predict any new particles at low scales.
 Furthermore, it is conceptually very close to
the idea proposed \cite{Einh} and used \cite{George} recently,
according to which each fermion
is associated with different extended-technicolor (ETC) scales,
instead of having a single technicolor scale for all of them,
as more conventional technicolor models suggest.
The difference here, of course, is that the hierarchy of ETC scales
is not introduced arbitrarily, but is produced by a specific
underlying mechanism.
Such a mechanism
would bring the chiral symmetry breaking scale in the picture
naturally, as
associated with the (common) technifermion self-energy at
low momenta. Then, it would automatically associate the lighter
 generations to the higher extended-technicolor scales.

 We also see that it is not
necessary  to assign a different ETC coupling to
 each fermion any more, as conventional technicolor models do,
  since this burdens the model with too many
  parameters. The change of the ETC-scales is enough to account
   for the change of the fermion masses from generation to generation.
 In addition, on can
 argue that we only have three generations, or
 equivalently that the solutions for $n > 3$ do not make sense,
because there are some new physics, above the scale
$\Lambda^{(3)}_{ETC}$, making our
analysis not applicable any more. This would give
 more predictive power to the proposed mechanism,
 since by using the known fermion spectrum we could
 have a feel of the order of magnitude where new
physics, beyond extended technicolor, enter into the picture.
 Note that
 such large energy scales ($~^{>}_{\sim}\; 10^{5}$ TeV) have been observed
 in highly-energetic cosmic rays \cite{Bird}.
These energies are still very far below
 a possible grand-unification scale or the Planck scale.

A  great advantage of such a mechanism is that it can avoid
 large flavor-changing neutral currents, since the ETC scale
  associated with the light quarks is high, while at the same
  time it can generate large bottom and top quark masses, since
   their are associated with a much smaller ETC scale.
   Of course, this also implies the existence of large FCNC
   associated with the third fermion generation,
   as well as non-negligible
   corrections to the $Z^{0} \rightarrow \bar{b}b$ vertex,
  effects that should be
   detectable in  precision experiments.
The smallness of the ETC scale associated with the top quark
has been shown to serve two more phenomenological purposes:
 it can keep small not only the S parameter, since
a ``walking" mechanism requiring many technifermions
is no longer needed to generate large
heavy quark masses, but also the $\Delta\rho$ parameter \cite{George}.

The problem with this mechanism is that it is theoretically
 unclear how each fermion generation is
 associated with each scale.
 Unlike the usual tumbling mechanism, where
 the scales introduced are the energies where the gauge
 interactions become so strong that they break the gauge group to
 a smaller one, the mechanism proposed here does not possess, at
 first sight at least, such a straightforward interpretation.
It would seem that it is only for specific ETC scales that the
Schwinger-Dyson equation can have non-trivial solutions and break
chiral symmetry. Then, it is this very dynamical symmetry breaking
that causes the ETC group to break
successively at these scales
  down to  smaller groups,
reproducing a mechanism similar to ``tumbling".
The correct physical interpretation of this
phenomenon is a very challenging model-building
  problem that we can hardly address here,
  and we will return to it, along with a more general
  physical discussion, in the next section.

We next go to the last remaining possibility, which is to fix
 $\Lambda_{ETC}$ and $\Sigma_{0}$, and to derive a quantization
 condition for $\delta$. The physical interpretation
 for such a picture  could be more straightforward than the
 previous one, since it would signal  that we
 have all the technifermion self-energies starting-off at
  low momenta from their
  common initial value $\Sigma_{0}$, and then drop up to
their common
 ETC scale according to different anomalous dimensions, i.e. with
different couplings. This could be very interesting from the
point of view of model-building, since
such a behavior could be attributed to having technifermions sitting in
different representations of the same technicolor group, or
having them interact with different technicolor groups altogether.
Unfortunately, since the relation between $\omega$ and
$\delta$ is non-trivial, especially in theories
where one  employs 4-fermion-induced high-momentum enhancement,
we do not pursue this analysis further, but
merely contend ourselves to stating
this interesting possibility.

\section { Conclusions }

In this work, we have attempted to construct
 a mechanism that would explain
the mass hierarchy of the three fermion generations, in
a context of dynamical electroweak and chiral symmetry breaking models.
We have tried to achieve this by using a minimum number of
input parameters, which makes these models more natural.

The explanation of the mass hierarchies in Nature
is however a highly non-trivial problem, and  attempts to solve it
usually give rise to serious complications. In our case, the solution
that is both solvable, after using several approximations, and
phenomenologically acceptable, is the one in which we fix
the weak scale, which is closely related to $\Sigma_{0}$, and
then the ETC scales follow from a quantization condition. Unfortunately,
such an interpretation is not along the lines of conventional wisdom
in present-day particle physics.  Let us see why this is so.

At first, we have to understand what the three fermion generations
correspond to in this picture. They
appear as the same reality that replicates itself, and manifests
itself into three different ways. In the everyday world, we can observe
only the lowest-energy manifestations of that reality, by means of the
lightest fermion generation. It is only when we go to higher energies that we
can see its higher-energy manifestations. The role of ETC scales, however, and
the exact way in which they enter in this process, is still unclear. They
appear as the scales which
can lead, when chiral symmetry breaking sets in,  to a
technifermion self-energy that is very close to
the weak scale at low momenta.

Moreover, what is usually expected in model building is for
high-energy physics to ``feed-down" their effects to
 lower energies.  The picture as presented here, on the contrary,
  seems to do the opposite;
 we first fix the weak scale and the coupling, and then we find the
 corresponding spectrum of ETC scales. It is as if lower-energy physics
 determine the behavior of higher-energy physics.

 This, however, is not a completely new phenomenon in the physical
 world. As a very
 naive and simplistic example, we  take the harmonic oscillator.
 One can completely define this quantum-mechanical system by
 specifying its fundamental frequency $\omega_{0}$. After solving
 the equations, however, we predict a whole spectrum of frequencies
 that are  arbitrarily larger than the fundamental one, in a
 similar way that our equations predict a spectrum of ETC scales
 much higher than the weak scale.
 Moreover, the reason for the appearance of a discrete spectrum
 in quantum mechanics is not always the existence of a bound state
 of two particles, but can also be the confinement of a particle in
 a finite space region.

 We are very much aware of the fact that analogies like
 the one above can lead to serious misconceptions.
 For instance, in our case the ETC scales are supposed to be
 physical cut-offs, and not the energy levels of a system of particles.
 The message that
 we want to convey should be clear nevertheless: we want to
 consider the weak scale, or equivalently the scale where new,
 strongly-interacting physics come into the picture at around 1 TeV,
 as a fundamental physical parameter which, by its inverse,
 sets a certain spatial scale.
 Within that finite space, the behavior of the Schwinger-Dyson equations
 generate a discrete spectrum of energy scales (cut-offs),
 which  could  possibly be
 identified with the ETC scales of technicolor theories.

 In addition, we should not
 forget that the fermion masses are much closer to the weak scale, rather
 than the Planck scale, so it does not seem to us too unnatural trying to
 explain them in terms of physics coming in at the weak scale, rather than
 expecting ``Planck-" or higher-scale
  physics to ``feed down" their effect directly to the fermion masses. Of
  course, the weak scale itself
  could still be determined by some unknown high-energy physics,
 appearing  at the ``Planck", or
  even at the highest ETC, scale.
  Therefore, the proposed mechanism, when
  seen from this point of view,
  does not  completely
  violate the way high-energy physics determine low-energy physics.
  It just gives the weak scale a more active and direct
  role in the fermion
  mass generation, while leaving for the Planck-scale, or for
  any other scale
  that determines the weak scale, only an indirect role.

 Such a picture still gives a very limited explanation of the mass hierarchies
 observed in nature. The huge hierarchy between the Planck scale and the weak
 scale still remains a mystery.
 Furthermore,
the QCD scale is another scale that is not accounted for in this picture.
Even though it is conceivable that these scales can be explained by a similar
paradigm, trying to incorporate them in the present discussion would be
over-ambitious.

To conclude, we would like to add the following comments.
The physical interpretation
of the proposed picture may still seem  elusive. In such
a case, it would still be interesting, as well as useful,
 to consider the formulas given here as
purely phenomenological, that merely describe,
which they seem to do indeed, instead of explaining, the true
situation. We should then await for a better understanding of the whole
process. Within the same framework, it would be also very useful to
perform a more detailed and careful mathematical analysis of the
quantization condition, and a more thorough investigation of
possible models and physical processes that could
explain the inner works of this mechanism.

\noindent {\bf Acknowledgements} \\
I thank Bob Holdom and Thomas Appelquist for helpful discussions.

\end{document}